\documentclass[twocolumn%
              ,superscriptaddress%
              ,aps%
              ,pra]{revtex4}
\usepackage[utf8]{inputenc}
\usepackage[T1]{fontenc}
\usepackage{amsmath,amssymb}
\usepackage{braket}
\usepackage{graphicx}
\usepackage{hyperref}

\newcommand{\e}{\mathrm{e}}
\newcommand{\tr}{\mathrm{tr}}
\newcommand{\TE}{\mathrm{TE}}
\newcommand{\TM}{\mathrm{TM}}
\newcommand{\E}{\mathrm{E}}
\newcommand{\M}{\mathrm{M}}
\newcommand{\out}{\mathrm{out}}
\newcommand{\reg}{\mathrm{reg}}
\newcommand{\Reff}{R_\mathrm{eff}}
\newcommand{\SP}{\mathrm{sp}}

\begin{document}
\title{Casimir effect between spherical objects: proximity-force approximation and beyond
       using plane waves}

\author{Tanja Schoger}
\affiliation{Institut für Physik, Universität Augsburg, 86135 Augsburg, Germany}

\author{Benjamin Spreng}
\affiliation{Department of Electrical and Computer Engineering, University of California, Davis, CA 95616, USA}

\author{Gert-Ludwig Ingold}
\affiliation{Institut für Physik, Universität Augsburg, 86135 Augsburg, Germany}

\author{Paulo A. Maia Neto}
\affiliation{Instituto de Física, Universidade Federal do Rio de Janeiro
Caixa Postal 68528, Rio de Janeiro, RJ, 21941-972, Brazil}

\begin{abstract}
For the Casimir interaction between two nearby objects, the plane-wave basis
proves convenient for numerical calculations as well as for
analytical considerations leading to an optical interpretation of the relevant
scattering processes of electromagnetic waves. We review work on the
proximity-force approximation and corrections to it within the plane-wave
basis for systems involving spherical objects. Previous work is extended by
allowing for polarization mixing during the reflection at a sphere. In particular,
explicit results are presented for perfect electromagnetic conductors.
Furthermore, for perfect electric conductors at zero temperature, it is demonstrated
that beyond the leading-order correction to the proximity-force approximation,
terms of half-integer order in the distance between the sphere surfaces appear.
\end{abstract}

\maketitle

\section{Introduction}

Even though Casimir's original calculation of the force between two objects due
to the electromagnetic vacuum considered two parallel plates
\cite{Casimir1948}, most experiments involve a sphere or a spherical lens,
often placed close to a plate (see, e.g.,
Refs.~\onlinecite{Klimchitskaya2006,Klimchitskaya2009,Decca2011,Lamoreaux2011,
Klimchitskaya2020, Mostepanenko2020, Gong2021} for reviews).  The advantage of
the sphere is that no special care needs to be taken to avoid misalignment.
Recently, experiments involving two spheres have been carried out as
well \cite{Ether2015,Garrett2018,Pires2021}. However, the strength of
the Casimir force is limited by the sphere radius. For most practical purposes,
the sphere radius therefore is chosen to be very large compared to the distance
to the other object. For the majority of experiments, the aspect ratio between
sphere radius and surface-to-surface distance lies between 100 and 5000
\cite{Hartmann2018a}. A notable exception is an experiment involving two
spheres with an aspect ratio smaller than 10 where optical tweezers are used in
order to measure a rather weak Casimir force \cite{Ether2015,Pires2021}.

For large aspect ratios, the force between the two involved objects can be
obtained to a very good approximation by dividing the opposing surfaces into
parallel surface elements and summing up the free energy for the respective
distances. This approximation was first introduced by Derjaguin
\cite{Derjaguin1934} and the term proximity force was coined by B{\l}ocki et
al.\ \cite{Blocki1977} leading to the often used term proximity-force
approximation (PFA). While the results of PFA are in many cases sufficient to
analyze experimental data, the increasing experimental precision \cite{Krause2007, Liu2019, Liu2019a} 
and the Drude-plasma controversy \cite{Decca2007, Mostepanenko2021} motivate to go beyond PFA. 
In special cases, it is possible to obtain analytical expressions for the leading order corrections to PFA, for instance
by developing a derivative expansion of the interaction energy~\cite{Fosco2011,Fosco2014}. 
However, in general one needs to resort to numerical techniques, particularly when dealing with moderate values for 
the aspect ratio or when a higher precision is required. 

There exist a number of numerical approaches to the Casimir effect, some of
them capable to handle rather general geometries \cite{Oskooi2010,Reid2015} and
others adapted to specific shapes of the involved objects.\cite{Hartmann2020}
Here, we will concentrate on setups containing two spheres or a sphere and a
plate.  Using a spherical wave basis may appear as quite natural and recently
it became indeed possible to push the limits of this method well into the
experimentally relevant regime \cite{Hartmann2017,Hartmann2018a}. It should be
kept in mind though that for a setup consisting of two spheres, two spherical
wave bases are actually needed, each of them centered at one of the spheres. As
a consequence, it is necessary to transform between these two bases.
Bispherical coordinates might appear as an alternative and they have in fact
been used to derive exact expressions for the Casimir force between two
spheres of equal radii as well as between a plane and a sphere in the high-temperature limit \cite{Bimonte2012a}.

More recently, it became clear that the use of a plane-wave basis can be
advantageous \cite{Spreng2020,Nunes2021,Spreng2021}, in particular when the aspect ratio
becomes large as is the case in most experiments. In view of the
proximity-force approximation where a more general geometry is locally
approximated by a plane-plane geometry, the advantages of the plane-wave basis
are comprehensible. An analysis of experimental data by numerical techniques
based on plane waves is presented in Ref.~\onlinecite{Bimonte2021}. The plane-wave
approach has also been employed to produce data for another article in this
volume \cite{Schoger2022}.

The plane-wave basis is not only useful for numerical computations but also in
analytical calculations. The proximity-force approximation for two spheres has
been derived by means of the scattering approach to the Casimir effect
\cite{Lambrecht2006,Emig2007,Ingold2015} as an asymptotic result using the
plane-wave basis \cite{Spreng2018}. Furthermore, extending this approach,
corrections to PFA have been obtained
\cite{Henning2019,Henning2021}. In the high-temperature limit, using the plane-wave basis, 
an exact expression for the Casimir free energy between two spheres of arbitrary radii was found
\cite{Schoger2021}. Some of the results had been at least partially
derived before by other means for the plane-sphere geometry at zero temperature \cite{Bordag2008,
Emig2008,MaiaNeto2008,Teo2011,Bimonte2012,Teo2013} and in the high-temperature limit
\cite{Canaguier-Durand2012} as well as for the sphere-sphere geometry at zero
temperature \cite{Teo2012} and for high temperatures \cite{Bimonte2012a,Fosco2016}.
On the other hand, particularly for large aspect ratios, a calculation based on plane
waves offers the opportunity for physical insights in terms of the concepts of
geometrical optics and its semiclassical corrections. 

In the following, we will discuss the Casimir interaction between spherical
objects in vacuum including the sphere-plane geometry as a limiting case from
the point of view of plane waves. While part of the paper will review previous
work, we will also present some new results. In particular, we will derive the
PFA expression in the plane-wave basis for polarization-mixing reflection at
the spheres, thereby generalizing previous work. Furthermore, we will point out
the appearance of so far unexpected corrections to the PFA result.

The paper is organized as follows. Section~\ref{sec:scattering_approach}
introduces the scattering approach to the Casimir free energy within the
plane-wave basis. The asymptotic expansion of the reflection coefficients
for large radii is developed in Sec.~\ref{sec:large_spheres}. In
Sec.~\ref{sec:asymptotic_expansion}, the Casimir free energy is evaluated in
the saddle-point approximation to obtain the PFA in the presence of 
polarization-mixing reflection. The leading-order correction 
to PFA is discussed in Sec.~\ref{sec:corrections_to_pfa}, where explicit
results are given for perfect electromagnetic conductors at zero temperature.  
Finally, Sec.~\ref{sec:NTLO_correction} explores the origin of the next-to-leading 
order correction to PFA. Concluding remarks are
given in Sec.~\ref{sec:conclusions}. 

\section{Scattering approach in the plane-wave basis} \label{sec:scattering_approach}

\subsection{General expression for the Casimir free energy}

When an object is placed into electromagnetic vacuum thereby acting as a scatterer, it will
give rise to a phase shift during the scattering process and, as a consequence, to a change
in the vacuum energy. If instead of a single object two objects are considered, the
Casimir energy is obtained by isolating the distance-dependent part of the vacuum energy.
The object describing the relevant part of the scattering process in the latter case is
the round-trip operator 
\begin{equation}
 \label{eq:roundtrip_operator}
 \mathcal{M} = \mathcal{R}_2\mathcal{T}_{21}\mathcal{R}_1\mathcal{T}_{12}
\end{equation}
containing reflections at the two objects described by the operators $\mathcal{R}_1$ and
$\mathcal{R}_2$ and translations between the two objects in terms of the operators
$\mathcal{T}_{21}$ and $\mathcal{T}_{12}$. 

The Casimir energy then reads
\begin{equation}
 \label{eq:casimir_energy_real}
 E = \frac{\hbar}{2\pi}\int_0^\infty d\omega \text{Im}\log\det(1-\mathcal{M}(\omega))\,.
\end{equation}
This expression is valid not only for the unitary case but also in the presence of dissipative
channels \cite{Guerout2018}. In order to avoid resonances on the real frequency axis, it is
convenient to perform a Wick transformation and to express the Casimir energy in terms of
imaginary frequencies $\xi = -i\omega$. Then, \eqref{eq:casimir_energy_real} becomes
\begin{equation}
 \label{eq:casimir_energy_imag}
 E = \frac{\hbar}{2\pi}\int_0^\infty d\xi \mathfrak{F}(\xi)
\end{equation}
with
\begin{equation}
 \label{eq:def_f_of_xi}
 \mathfrak{F}(\xi) = \log\det\big(1-\mathcal{M}(\xi)\big)\,.
\end{equation}
In the following, we will always make use of imaginary frequencies unless stated otherwise.
At finite temperatures, thermal fluctuations of the electromagnetic field have to be accounted
for as well. This can be done within the Matsubara formalism where the Casimir free energy at
temperature $T$ is found as
\begin{equation}
 \label{eq:casimir_free_energy}
 \mathcal{F} = \frac{k_\mathrm{B}T}{2}\sum_{n=-\infty}^{\infty}\mathfrak{F}(\vert\xi_n\vert)
\end{equation}
with the Matsubara frequencies $\xi_n = 2\pi n k_\mathrm{B}T/\hbar$.
Making use of the mathematical identity $\log\det(1-\mathcal{M}) = \text{tr}\log(1-\mathcal{M})$
we can expand the logarithm and decompose the free energy into contributions of $r$ round-trips
between the two scatterers as
\begin{equation}
 \label{eq:roundtrip_decomposition}
 \mathfrak{F}(\xi) = -\sum_{r=1}^\infty\frac{\text{tr}\mathcal{M}^r(\xi)}{r}\,.
\end{equation}
The evaluation of the trace requires the choice of an appropriate basis. 
While spherical waves are practical for distances large compared to the diameter of 
the objects, plane waves turn out to be more suitable for short distances \cite{Spreng2018}
and will be chosen in the following.

We characterize the plane waves in our basis by its wave vector and polarization. It is 
convenient to decompose the wave vector into its projection $\mathbf{k}$ onto the plane
perpendicular to the axis connecting the two scattering objects. The wave vector component
$\kappa$ along the axis is taken imaginary as we did with the frequency $\xi$ so that the
dispersion relation becomes
\begin{equation}
 \label{eq:dispersion_relation}
 \kappa = \left(\frac{\xi^2}{c^2}+k^2\right)^{1/2}
\end{equation}
with the modulus $k$ of the transverse wave vector $\mathbf{k}$. Since $\kappa$
is positive, we denote the sense of propagation by a sign $\phi=\pm$. 

\begin{figure}
 \begin{center}
  \includegraphics[width=0.8\columnwidth]{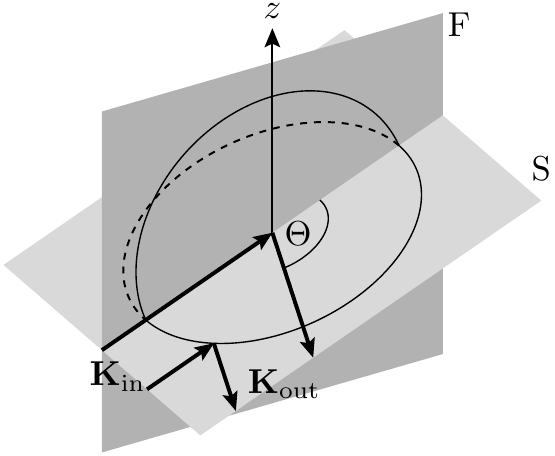}
 \end{center}
 \caption{The Fresnel plane (F) is spanned by the $z$-axis and the incident wave vector
	  $\mathbf{K}_\text{in}$ while the scattering plane (S) is spanned by the incident
	  wave vector and the wave vector $\mathbf{K}_\text{out}$ of the reflected wave.
	  The scattering angle is denoted by $\Theta$.}
 \label{fig:fresnel_scattering_plane}
\end{figure}

The polarization $p$ of the plane wave is defined with respect to the Fresnel plane (F). As
shown in Fig.~\ref{fig:fresnel_scattering_plane}, the Fresnel plane is spanned
by the axis connecting the two scattering objects, \textit{i.e.} the $z$-axis in the figure,
and the incident real wave vector $\mathbf{K}_\text{in}$. We distinguish between transverse magnetic
($p=\TM$) and transverse electric ($p=\TE$) modes, where the electric field is parallel or
perpendicular to the Fresnel plane, respectively.

An incident plane wave of our basis is denoted in the angular spectral representation
\cite{NietoVesperinas2006} as $\vert\mathbf{k}, p, \phi\rangle$. Here, we omit the imaginary
frequency $\xi$ which is conserved through the complete scattering process. As $\kappa$
is defined through the dispersion relation \eqref{eq:dispersion_relation}, we do not need
to include it in our notation. The specification of the plane wave is completed by the
polarization $p$ and the propagation direction indicated by the sign $\phi=\pm$.

We are now in a position to explicitly express the trace appearing in the round-trip expansion
\eqref{eq:roundtrip_decomposition} of the free energy in terms of the plane-wave basis.
The trace over the $r$-th power of the round-trip operator reads
\begin{equation}
\begin{aligned}
\label{eq:def_trMr}
\mathrm{tr}\mathcal{M}^r &= \sum_{p_1,\ldots, p_{2r}} 
\int\frac{\mathrm{d}\mathbf{k}_1\ldots 
\mathrm{d}\mathbf{k}_{2r}}{(2\pi)^{4r}} 
\prod_{j=1}^{r} \e^{-\kappa_{2j}\mathcal{L}} \e^{-\kappa_{2j-1}\mathcal{L}}\\
&\quad
\times \langle \mathbf{k}_{2j+1}, p_{2j+1}, - \vert
\mathcal{R}_2 \vert \mathbf{k}_{2j}, p_{2j}, + \rangle \\
&\quad\times
\langle \mathbf{k}_{2j}, p_{2j}, + \vert
\mathcal{R}_1 \vert \mathbf{k}_{2j-1}, p_{2j-1}, - \rangle\,,
\end{aligned}
\end{equation}
where cyclic indices $2r+1 \equiv 1$ were introduced to account for the trace. The product
in \eqref{eq:def_trMr} contains exponential factors arising from the translation operators
$\mathcal{T}_{ij}$ in \eqref{eq:roundtrip_operator} which in the plane-wave basis are
diagonal and describe the translation over the distance $\mathcal{L}$ between the two
scattering objects. The reflection matrix elements will in general be nondiagonal and 
depend on the details of the objects scattering the electromagnetic waves. So far, expression
\eqref{eq:def_trMr} is general. In the next section, we specialize to spherical objects
which will allow us to obtain explicit expressions for the reflection matrix elements.

\subsection{Reflection matrix elements}

The scattering of electromagnetic waves at a sphere can be solved analytically \cite{Mie1908}
and Mie scattering is discussed in various textbooks \cite{vandeHulst1981,BohrenHuffman2004}.
We therefore concentrate in this section only on aspects relevant for our purposes.

In the spherical-wave basis characterized by the angular momentum eigenvalues $\ell$ and $m$
as well as the polarization $P=\text{E}, \text{M}$, which can be electric or magnetic, respectively,
the matrix elements of the reflection operator are given by
\begin{equation}
\label{eq:generalized_mie}
\langle\ell', m', P', \out\vert \mathcal{R}\vert \ell, m, P, \reg\rangle = 
r_\ell^{P' P}(i\tilde{\xi}) \delta_{\ell,\ell'} \delta_{m, m'}
\end{equation}
with the size parameter 
\begin{equation}
 \label{eq:size_parameter}
 \tilde{\xi} = \frac{\xi R}{c}\,.
\end{equation}
Here, $R$ is the sphere radius and $c$ denotes the speed of light. We use `reg' to refer to a spherical
wave which is regular at the sphere center while `out' refers to an outgoing spherical wave \cite{vandeHulst1981}.

Due to the spherical geometry, $\ell$ and $m$ are conserved during the scattering process.
While mostly isotropic materials are discussed in the literature, where also the polarization
is conserved, we will allow here for the more general case where the polarization may change
as a result of the scattering at the sphere. Such a situation is encountered for spheres made
of a bi-isotropic material characterized by the constitutive relations
\cite{Sihvola1991}
\begin{equation}
\begin{aligned}
 \label{eq:constitutive_relations}
 \mathbf{D} &= \varepsilon \mathbf{E} + \alpha \mathbf{H} \\
 \mathbf{B} &= \mu \mathbf{H} + \beta \mathbf{E} 
\end{aligned}
\end{equation}
relating the electric displacement $\mathbf{D}$ and the magnetic induction $\mathbf{B}$ to
the electric field $\mathbf{E}$ and the magnetic field $\mathbf{H}$. Among the quantities
characterizing the material, $\varepsilon$ and $\mu$ are assumed to be scalars while
$\alpha$ and $\beta$ are pseudoscalars. An important class of materials described by such
relations are chiral materials \cite{Condon1937} where $\alpha=-\beta$.
Taking the prefactors to infinity, perfect
electromagnetic conductors (PEMC) \cite{Lindell2005} interpolating between perfect electric
conductors ($\varepsilon\to\infty$) and perfect magnetic conductors ($\mu\to\infty$) are also
covered by \eqref{eq:constitutive_relations}, if one takes $\alpha = \beta$. PEMC have recently 
been discussed in the context of the Casimir effect \cite{Rode2018}.

As already pointed out above, we want to make use of the plane-wave basis. The
Mie scattering amplitudes $S_{p'p}$ describe the scattering of a plane wave
with polarization $p$ by the sphere into a plane wave with polarization $p'$.
The scattering geometry is defined by the in- and outgoing wave vectors
spanning the scattering plane indicated by S in
Fig.~\ref{fig:fresnel_scattering_plane}. It is also with respect to this plane
that the polarization is defined. The two wave vectors define a scattering
angle $\Theta$ also shown in Fig.~\ref{fig:fresnel_scattering_plane}. For the
imaginary frequency $\xi$ and wave-vector component $\kappa$ introduced earlier, the
scattering angle is determined through
\begin{equation}
 \label{eq:cos_theta}
 \cos(\Theta)  = -\frac{c^2}{\xi^2}(\mathbf{k}_\text{out}\cdot\mathbf{k}_\text{in}
				    + \kappa_\text{out}\kappa_\text{in})\,.
\end{equation}
Explicit expressions for the Mie scattering amplitudes have been derived in Ref.~\onlinecite{Bohren1974}.
With our notation for the reflection matrix elements \eqref{eq:generalized_mie},
they are given by 
\begin{align}
\label{eq:def_S1}
S_{\TE,\TE}(\Theta) &= -\sum_{\ell=1}^\infty \frac{2\ell+1}{\ell(\ell+1)} 
	( \tau_\ell r_\ell^{\M\M}  + \pi_\ell r_\ell^{\E\E} )\\
\label{eq:def_S2}
S_{\TM,\TM}(\Theta) &= -\sum_{\ell=1}^\infty \frac{2\ell+1}{\ell(\ell+1)} 
	(\tau_\ell r_\ell^{\E\E}  + \pi_\ell r_\ell^{\M\M})\\ 
\label{eq:def_S3}
S_{\TM,\TE}(\Theta) &= i\sum_{\ell=1}^\infty \frac{2\ell+1}{\ell(\ell+1)} 
	(\tau_\ell r_\ell^{\E\M}  + \pi_\ell r_\ell^{\M\E})\\
\label{eq:def_S4}
S_{\TE,\TM}(\Theta) &= -i\sum_{\ell=1}^\infty \frac{2\ell+1}{\ell(\ell+1)} 
	(\tau_\ell r_\ell^{\M\E} + \pi_\ell r_\ell^{\E\M})\,.
\end{align}
While in the usual case of isotropic spheres only the first two scattering
amplitudes are non-vanishing, we need to account for all four of them
as we are dealing with bi-isotropic spheres. The scattering amplitudes not only
depend on the reflection matrix elements \eqref{eq:generalized_mie} but also
contain the scattering geometry through the angular functions \cite{BohrenHuffman2004}
\begin{align}
 \label{eq:angular_function_pi}
 \pi_\ell &= \frac{P_\ell^1\big(\cos(\Theta)\big)}{\sin(\Theta)} \\
 \label{eq:angular_function_tau}
 \tau_\ell &= \frac{d P_\ell^1\big(\cos(\Theta)\big)}{d\Theta}\,,
\end{align}
where $P_\ell^1$ are associated Legendre polynomials \cite{DLMF}.

So far, the scattering plane has been used in the calculations. However, for
our purposes in a two-sphere setup, the Fresnel plane is more convenient. In a 
last step, we thus have to transform the results from the scattering plane to the
Fresnel plane indicated as S and F, respectively, in
Fig.~\ref{fig:fresnel_scattering_plane}. The matrix elements needed for this
basis change can be found in Refs.~\onlinecite{Mishchenko2000} and \onlinecite{Messina2015}. Finally,
one arrives at the reflection matrix element in the plane-wave basis
\begin{equation}
\begin{aligned}
\label{eq:reflection_matrix_elements}
 \langle\mathbf{k}_j, p_j\vert \mathcal{R}\vert \mathbf{k}_i, p_i\rangle &=
 \frac{2\pi c}{\xi \kappa_j}\big[A S_{p_jp_i}
	+(-1)^{p_j+p_i} B S_{\bar{p}_j\bar{p}_i}\\
	&\quad- (-1)^{p_j} C S_{\bar{p}_jp_i}
	+ (-1)^{p_i} D S_{p_j\bar{p}_i}\big]\,,
\end{aligned}
\end{equation}
where the polarization of the in- and outgoing plane waves is defined with
respect to the Fresnel plane and the transverse wave vector is taken with
respect to the axis connecting the two spheres. In the exponents, $p$ takes the
values $1$ and $2$ for polarizations TE and TM, respectively, and the bar over
a polarization refers to the other polarization, \textit{i.e.} $\bar p=\TM$ if
$p=\TE$ and vice versa. The coefficients $A, B, C,$ and $D$ are functions of
the three-dimensional wave vectors, \textit{e.g.} $A = A(\mathbf{K}_j,
\mathbf{K}_i)$.  Since we will not need the complete expression of these
coefficients in the following, we refer the reader interested in more details
to Ref.~\onlinecite{Spreng2018} and in particular to appendix~A of that paper for
explicit expressions for the coefficients $A, B, C,$ and $D$.

\section{Scattering at large spheres}\label{sec:large_spheres}
The proximity-force approximation can be obtained as leading-order term in an asymptotic
expansion for large sphere radii $R_1$ and $R_2$ and corrections are given by higher-order
terms. An important ingredient are the reflection matrix elements for a sphere which we
will expand in this section for large radii $R$. For real frequencies, the leading-order
term of the reflection matrix elements corresponds to the geometrical optics limit which
was extensively discussed \textit{e.g.} in Refs.~\onlinecite{Nussenzveig1969, vandeHulst1981, 
Nussenzveig1992, Grandy2000}.
In order to evaluate corrections to the proximity-force approximation, we will also need
to consider at least the leading correction to the reflection matrix elements which gives
rise to diffractive corrections. As a consequence, while the PFA result can be obtained
within geometrical optics, corrections to PFA will contain contributions from geometrical
optics as well as from diffraction.

When expanding the reflection matrix elements, we will need to assume that the radius $R$
is large compared to other length scales of the problem, which in our case is the wave number
$\xi/c$. While for non-zero values of $\xi$ it is sufficient to take the limit of a large
size parameter \eqref{eq:size_parameter}, we have to consider the case $\xi=0$ separately.

\subsection{Reflection coefficients for finite frequencies}

In the expansion of the scattering amplitudes
\eqref{eq:def_S1}--\eqref{eq:def_S4} appearing in
\eqref{eq:reflection_matrix_elements} for large sphere radius, the localization
principle \cite{vandeHulst1981} plays an important role. According to this
principle, the scattering of a ray with an impact parameter $b$ is dominated by
angular momenta of the order of $\xi b/c$. Applying Debye's expansion \cite{DLMF}
to the reflection coefficients $r_\ell^{P'P}$ defined through
\eqref{eq:generalized_mie}, the asymptotic expansion in the
multipole basis is already completed. The resulting expressions for the Mie
coefficients in the isotropic case have been used \textit{e.g.} in
Ref.~\onlinecite{Teo2013} to obtain the leading correction to the proximity-force
approximation in the plane-sphere geometry.

\begin{figure}
 \begin{center}
  \includegraphics[width=0.8\columnwidth]{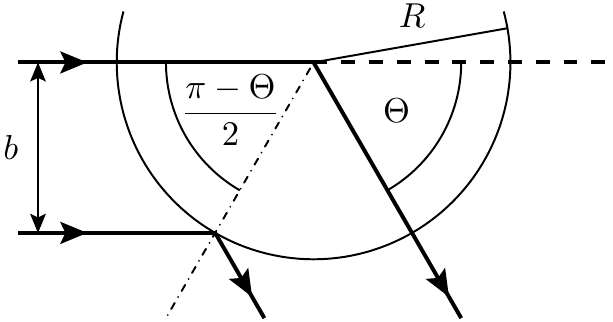}
 \end{center}
 \caption{Scattering geometry at a sphere of radius $R$ for a scattering angle $\Theta$ (cf.\
	  Fig.~\ref{fig:fresnel_scattering_plane}). The corresponding impact parameter $b$ in
	  ray optics is given by $R\cos(\Theta/2)$.}
 \label{fig:ray_scattering}
\end{figure}

Within the plane-wave basis, we need to go one step further and evaluate the sum
over the angular momenta in \eqref{eq:def_S1}--\eqref{eq:def_S4}. Since the
procedure for real frequencies is well known from textbooks and can be carried
over to imaginary frequencies without any difficulties, we will restrict
ourselves to outlining the main ideas. In view of the localization principle,
we need to account for angular momenta $\lambda = \ell + 1/2 \lesssim
\tilde\xi$. Since the size parameter $\tilde\xi$ defined in
\eqref{eq:size_parameter} for a fixed frequency $\xi$ becomes large for large
radius $R$, we can approximate the sums in \eqref{eq:def_S1}--\eqref{eq:def_S4}
by integrals over $\lambda$. In addition, an asymptotic expansion of the angular
functions \eqref{eq:angular_function_pi} and \eqref{eq:angular_function_tau} for large orders $\ell$ is used.
Then, the dominant contribution to the integral can be obtained by means of the
saddle-point approximation with the saddle point given by
\begin{equation}
 \lambda_\text{sp} = i\tilde{\xi}\cos\left(\frac{\Theta}{2}\right)\,.
\end{equation}
For real frequencies, the saddle point corresponds to a ray hitting the sphere
with an impact parameter $b = \lambda_\text{sp} c/i\xi = R\cos(\Theta/2)$ as depicted in
Fig.~\ref{fig:ray_scattering}, where $\Theta$ is the scattering angle. 

After proceeding as for real frequencies \cite{Grandy2000}, one finds for the scattering
amplitudes for imaginary frequencies
\begin{equation}
\label{eq:WKB_scattering_amplitude}
S_{p'p}(\Theta) = \frac{\tilde{\xi}}{2} 
e^{2\tilde{\xi}\sin(\Theta/2)}
\tilde{r}_{p'p}(\Theta)
\end{equation}
with 
\begin{equation}
\label{eq:ref_coef_corr}
\tilde{r}_{p'p}(\Theta) =  
r_{p'p}\left(\frac{\pi-\Theta}{2}\right)
\left[ 1 + \frac{s_{p'p}}{\tilde{\xi}} + \mathcal{O}\left(\tilde{\xi}^{-2}\right)\right]\,.
\end{equation}
The exponential term in \eqref{eq:WKB_scattering_amplitude} accounts for the difference in
path lengths between a ray reflected at the sphere surface and a ray going to the sphere origin
and being deflected there by the scattering angle $\Theta$ as depicted in Fig.~\ref{fig:ray_scattering}.
The latter ray is relevant for us because the origin of the sphere was chosen as origin of
the reference frame for the plane-wave basis.

In \eqref{eq:ref_coef_corr}, $r_{p'p}$ are the reflection coefficients at a plane evaluated
at an angle of incidence $(\pi - \Theta)/2$ corresponding to a scattering angle $\Theta$ as can
be seen from Fig.~\ref{fig:ray_scattering}. The leading-order term for large radii then does not
depend on the curvature of the sphere. In the absence of polarization mixing, the reflection
coefficients correspond to the well-known Fresnel reflection coefficients \cite{BohrenHuffman2004}.
For bi-isotropic materials, the reflection coefficients can be found in Ref.~\onlinecite{Lindell1994}.
The leading corrections are given by $s_{p'p}$ and explicit expression have been determined in
Refs.~\onlinecite{Grandy2000} and \onlinecite{Khare1975} for isotropic materials. However, the expressions
given in the two references do not agree and are also inconsistent with numerical results. 
Therefore, for reference, we give the explicit expressions derived in Ref.~\onlinecite{Spreng2021}
which were checked against the numerical results. Introducing the abbreviations $s\equiv\sin(\Theta/2)$
and $c\equiv\cos(\Theta/2)$, the corrections for a dielectric nonmagnetic material with index of
refraction $n = \epsilon^{1/2}$ read
\begin{equation}
 \begin{aligned}
 \label{eq:sTETE}
 s_{\TE,\TE} &= \frac{1-2s^2}{2s^3} + \frac{1}{s}\frac{1}{c^2+s(n^2-c^2)^{1/2}}\\
	     &\quad -\frac{2n^2-c^2}{2(n^2-c^2)^{3/2}}\\
 \end{aligned}
\end{equation}
\begin{equation}
 \begin{aligned}
 s_{\TM,\TM} &= -\frac{1}{2s^3} + \frac{1}{s}\frac{1}{c^2-s(n^2-c^2)^{1/2}}\\
	     &\quad -\frac{c^2}{s^3}\frac{2n^4s^2-n^2c^2(1+s^2-s^4)+c^6}{(n^2-c^2)(n^2s^2-c^2)^2}\\
	     &\quad +\frac{n^2}{2(n^2-c^2)^{3/2}}\frac{2n^4-n^2c^2(1+c^2)-c^4}{(n^2s^2-c^2)^2}\,.\label{eq:sTMTM}
 \end{aligned}
\end{equation}
For perfect reflectors, which we will always assume to be made of a perfectly electric conductor,
$n\to\infty$ and all terms except for the first one in each case vanish and the results agree
with Refs.~\onlinecite{Grandy2000} and \onlinecite{Khare1975} in that limit.

As a definite example where polarization mixing occurs during a reflection, we consider
a sphere made of a perfect electromagnetic conductor (PEMC) \cite{Ruppin2006}, a model introduced
above. PEMC are characterized by an angle $\theta$ (not to be confused with the scattering angle $\Theta$)
which describes the transition from a perfect electric conductor with $\theta=0$ to a 
perfect magnetic conductor with $\theta=\pi/2$. The reflection coefficients including the
first correction in $1/\tilde\xi$ are given by
\begin{equation}
 \begin{aligned}
\label{eq:rTETE_pemc}
 \tilde r_{\TE,\TE} &= -\cos(2\theta)\\
	            &\quad- \frac{1}{\tilde\xi}\left(\frac{1-2s^2}{2s^3}\cos(2\theta)
			  +\frac{c^2}{s^3}\sin^2(\theta)\right) + O(\tilde\xi^{-2})
 \end{aligned}
\end{equation}
\begin{equation}
 \begin{aligned}
 \tilde r_{\TM,\TM} &= \cos(2\theta)\\
	            &\quad + \frac{1}{\tilde\xi}\left(\frac{1-2s^2}{2s^3}\cos(2\theta)
			  -\frac{c^2}{s^3}\cos^2(\theta)\right) + O(\tilde\xi^{-2})
 \end{aligned}
\end{equation}
\begin{align}
 \tilde r_{\TE,\TM} &= \tilde r_{\TM,\TE}
                     = -\sin(2\theta) + \frac{1}{\tilde\xi}\frac{1}{2s}\sin(2\theta) + O(\tilde\xi^{-2})\,,
\label{eq:rTETM_pemc}                  
\end{align}
where the leading term corresponds to the reflection at a plane PEMC surface \cite{Rode2018}.

With the expansion of the scattering amplitude for large radii \eqref{eq:WKB_scattering_amplitude},
the reflection matrix elements \eqref{eq:reflection_matrix_elements} appearing in the trace over
the $r$-th power of the round-trip operator \eqref{eq:def_trMr} are given by
\begin{equation}
\label{eq:WKB_approximation}
\langle\mathbf{k}_j, p_j, \pm\vert \mathcal{R}\vert \mathbf{k}_i, p_i, \mp \rangle 
= \frac{\pi R} {\kappa_j}e^{2\tilde{\xi}\sin(\Theta/2)}
\rho_{p_j p_i}\,,
\end{equation}
where
\begin{equation}
 \begin{aligned}
\label{eq:rho_pp'}
\rho_{p_jp_i} &= A \tilde{r}_{p_j p_i} 
	+(-1)^{p_j+p_i} B \tilde{r}_{\bar{p}_j\bar{p}_i}
	-(-1)^{p_j}C \tilde{r}_{\bar{p}_jp_i}\\ 
	 &\quad + (-1)^{p_i}D \tilde{r}_{p_j\bar{p}_i}\,.
 \end{aligned}
\end{equation}
For the discussion in Sec.~\ref{sec:asymptotic_expansion} it will be relevant
that in view of the definition of the size parameter \eqref{eq:size_parameter},
the factorization in the expression \eqref{eq:WKB_approximation} for the
reflection matrix element separates an exponential dependence on the sphere
radius from a non-exponential dependence via the factors $\rho_{p_jp_i}$.

\subsection{Reflection coefficients at zero frequency}
\label{eq:refl_coeff_zero}

The previous section assumed a size parameter $\tilde\xi\gg1$ which at room temperature is
fulfilled for all non-vanishing Matsubara frequencies provided that ${R\gg 1.2\,\mu\text{m}}$.
However, for the zero Matsubara frequency $\xi_0$ this condition for the size parameter cannot
be fulfilled so that this case needs to be considered separately. In contrast to the previous
subsection, the reflections coefficients and the angular functions need to be evaluated in the
low-frequency limit.

Details of the zero-frequency case are given in appendix~B of Ref.~\onlinecite{Spreng2018}, so that again
we only highlight a few important points. According to \eqref{eq:cos_theta}, the cosine of the
scattering angle $\Theta$ diverges like $\xi^{-2}$ in the low-frequency limit. As a consequence,
the angular functions \eqref{eq:angular_function_pi} and \eqref{eq:angular_function_tau} diverge
as well. One finds that $\tau_\ell\sim\xi^{-2\ell}$ diverges more strongly than $\pi_\ell$ so that
the latter angular function can be disregarded. On the other hand, the reflection coefficients
vanish for $\xi \rightarrow 0$ with the explicit scaling depending on the response functions
of the sphere and the surrounding medium. Reflection coefficients in the multipole basis are
found to lead to a relevant contribution if they scale like $\xi^{2\ell+1}$ for angular momentum
$\ell$. Then, the scattering amplitudes are proportional to $\xi$ for small frequencies as is
the case in the finite-frequency result \eqref{eq:WKB_scattering_amplitude}. 

For small frequencies, the scattering amplitudes can be expressed as
\begin{equation}
\label{eq:S_freq0}
S_{p'p} = \tilde{\xi} \sum_{\ell = 1}^\infty 
X_{p'p}(\ell)\frac{[-2\tilde{\xi}^2\cos(\Theta)]^{\ell}}{(2\ell)!}
\end{equation}
with $X_{p'p}(\ell)$ depending on the materials considered. In Ref.~\onlinecite{Spreng2018}
expressions for dielectrics in vacuum as well as for perfect electric conductors are given. The
zero-frequency result for another situation involving two dielectric spheres in an electrolyte
is discussed in another article in the present volume \cite{Schoger2022}.

For the special case of a PEMC sphere, the model parameters $X_{p'p}(\ell)$ are given by
\begin{align}
\label{eq:X_TETE}
X_{\TE, \TE}(\ell) &= \sin^2(\theta) \mathcal{A}_\ell^\text{PEC} + 
	\cos^2(\theta) \mathcal{B}_\ell^\text{PEC} \\
X_{\TM, \TM}(\ell) &= 	\cos^2(\theta) \mathcal{A}_\ell^\text{PEC} +
	\sin^2(\theta) \mathcal{B}_\ell^\text{PEC} \\
X_{\TE, \TM}(\ell) &= X_{\TM, \TE}(\ell)\nonumber\\
       &=-\frac{1}{2}\sin(2\theta)\left[\mathcal{A}_\ell^\text{PEC}
	 -\mathcal{B}_\ell^\text{PEC}\right]\,,
\label{eq:X_TETM}	 
\end{align}
where
\begin{equation}
\mathcal{A}_\ell^\text{PEC} = 1, \quad 
\mathcal{B}_\ell^\text{PEC} = -\frac{\ell}{\ell+1}
\end{equation}
correspond to the model parameters of the $\TM$ and $\TE$ modes, respectively, for a perfect electric
conductor (PEC).  

As mentioned above $\cos(\Theta) \sim \xi^{-2}$, so that the numerator in \eqref{eq:S_freq0} is large 
for large spheres and the main contribution again comes from large angular momenta $\ell$. Replacing
the factorial by Stirling's approximation and applying the saddle-point approximation one finds a
saddle point at
\begin{equation}
 \ell_\text{sp} = \tilde{\xi}\sin\left(\Theta/2\right)\,. 
\end{equation}

At zero frequency, the transformation from the scattering plane to the Fresnel
plane does not lead to a polarization change so that the coefficients in
\eqref{eq:reflection_matrix_elements} become $A=1$ and $B=C=D=0$. Therefore,
the reflection matrix elements at zero frequency are given by
\eqref{eq:WKB_approximation} if we replace $\rho_{p_j p_i}$ by $X_{p_j
p_i}(\ell_\SP)$.

For dielectrics and perfect electric conductors it was shown in
Ref.~\onlinecite{Spreng2021} that the leading order of $X_{p_j p_i}(\ell_\SP)$
agrees with the reflection coefficients $r_{p_j p_i}$ in
\eqref{eq:ref_coef_corr}. This equivalence can be generalized to bi-isotropic
spheres provided the material parameters given in
\eqref{eq:constitutive_relations} are finite \cite{SchogerUnpub} and it can be seen to hold also
for the special case of PEMC spheres with the model parameters given by
\eqref{eq:X_TETE}--\eqref{eq:X_TETM}.

While the leading order of the reflection matrix elements \eqref{eq:reflection_matrix_elements} for finite and
zero frequency coincide, this is not true for the leading corrections. Here, the
full expression of $X_{p_j p_i}(\ell_\SP)$ needs to be taken into account.
Thus, the PFA result derived in Sec.~\ref{sec:asymptotic_expansion} holds for
arbitrary temperature. On the other hand, we limit our discussion of the
leading corrections to PFA in Secs.~\ref{sec:corrections_to_pfa} and
\ref{sec:NTLO_correction} to the case of zero temperature while the treatment
of finite temperatures is beyond the scope of this paper.

\section{Proximity-force approximation in the presence of polarization mixing}
\label{sec:asymptotic_expansion}

Based on the large-sphere approximation of the reflection matrix elements
\eqref{eq:WKB_approximation}, the trace \eqref{eq:def_trMr} over a power of
round-trip operators can be evaluated within the saddle-point approximation
(SPA). As discussed in detail in Ref.~\onlinecite{Spreng2018} for dielectric
spheres, the lowest order of the SPA leads to the proximity-force approximation
of the Casimir interaction. Here, we will generalize this result by allowing
for polarization mixing during the reflection at both spheres. We review the
main steps of Ref.~\onlinecite{Spreng2018} and go into more detail where the effects
of polarization mixing become relevant. In Sec.~\ref{sec:corrections_to_pfa},
we will then address the leading corrections which arise from higher-order
terms of the SPA as well as the first correction appearing in the reflection
coefficients \eqref{eq:ref_coef_corr}.

For the saddle-point approximation, we need to factor the integrand  in
\eqref{eq:def_trMr} into a term depending exponentially on the sphere radii and
a remaining term as
\begin{align}
\label{eq:trMr_WKB}
\tr \mathcal{M}^r \approx \left(\frac{R_1R_2}{16\pi^2}\right)^r 
\int \mathrm{d}^{2r} \mathbf{k} \,
& g(\mathbf{k}_1, \ldots, \mathbf{k}_{2r}) 
e^{- f(\mathbf{k}_1, \ldots, \mathbf{k}_{2r})}\,.
\end{align}
The function in the exponent is given by 
\begin{equation}
\label{eq:def_f}
f(\mathbf{k}_1, \ldots, \mathbf{k}_{2r}) = \sum_{j=1}^r 
(R_1 \eta_{2j} + R_2 \eta_{2j-1})\,,
\end{equation}
where $R_1$ and $R_2$ are the radii of the two spheres and
\begin{equation}
\label{eq:def_eta}
\eta_i = \kappa_i + \kappa_{i+1}
- \left[2\left(\frac{\xi^2}{c^2} + \kappa_i \kappa_{i+1} + 
\mathbf{k}_i \cdot \mathbf{k}_{i+1}\right)\right]^{1/2}\,.
\end{equation}
The first two terms in $\eta_i$ arise from the translation operators, specifically
from the parts of the translation within the spheres, while the last term arises
from the exponential term in the reflection matrix element \eqref{eq:WKB_approximation}
corresponding to the phase acquired upon scattering at the sphere as discussed in
connection with Fig.~\ref{fig:ray_scattering}. The function $g$ collects the remaining
terms and is given by 
\begin{equation}
 \begin{aligned}
 \label{eq:def_g}
g(\mathbf{k}_1, \ldots, \mathbf{k}_{2r}) &= 
\sum_{p_1,\ldots, p_{2r}} \prod_{j=1}^r 
\frac{e^{-(\kappa_{2j} + \kappa_{2j-1})L}}{\kappa_{2j} \kappa_{2j-1}}\\
	 &\qquad\qquad\qquad\times\rho^{(1)}_{p_{2j+1} p_{2j}}
\rho^{(2)}_{p_{2j}p_{2j-1}}\,.
 \end{aligned}
\end{equation}
Here, the remaining part of the contribution of the translation operator accounts for the
closest surface-to-surface distance $L$ which is assumed to be much smaller than the sphere
radii $R_1$ and $R_2$.

We will first calculate the leading-order saddle-point approximation (LO-SPA) of the integral
\eqref{eq:trMr_WKB} for the general case of bi-isotropic spheres. Explicit expressions are given
at the end of this subsection for PEMC spheres. 

The gradient of the function $f$ defined in \eqref{eq:def_f} vanishes for
\begin{equation}
\label{eq:saddle_point}
\mathbf{k}_1 = \ldots =\mathbf{k}_{2r} \equiv \mathbf{k}_\SP\,,
\end{equation}
thus leading to a two-dimensional manifold of saddle points (sp) over which one needs to integrate over.
The physical implications of these saddle points will be discussed below.

The Hessian matrix $\mathsf{H}$ governing the deviations from the saddle point manifold has block-diagonal form,
if ordered according to the $x$- and $y$-component of the transverse wave vector. The elements of each block, 
evaluated on the saddle-point manifold, are given by blocks of $2r$-dimensional
circulant matrices
\begin{equation}
\begin{aligned}
\left(\mathsf{H}_{\alpha, \alpha}\right)_{ij} 
 &= \frac{\partial^2 f}{\partial k_{i, \alpha}\partial k_{j, \alpha}} \Bigg|_{\SP}\\
 &= \frac{1}{2\kappa_\SP}\big[(R_1 + R_2)\delta_{i,j} - R_{2-(i+1)\,\text{mod}2}\bar\delta_{i+1,j} \\
 &\qquad\qquad - R_{2-i\,\text{mod}2}\bar\delta_{i, j+1}\big]
\end{aligned}
\end{equation}
with $\alpha = x, y$ and $\bar\delta$ denoting a Kronecker delta symbol where the
equality of indices is taken modulo $2r$.

Carrying out the saddle-point approximation, one finds
\begin{equation}
\label{eq:trMr_LO-SPA}
\left[\tr\mathcal{M}^r\right]_\text{LO-SPA}  
=2r\left(\frac{R_1R_2}{4}\right)^r \int \frac{\mathrm{d}\mathbf{k}_\SP}{2\pi} 
\frac{e^{-f_\SP}}{\sqrt{\text{pdet}(\mathsf{H})}}g_\SP\,,
\end{equation}
where $f_\SP$ and $g_\SP$ are given by \eqref{eq:def_f} and \eqref{eq:def_g}, respectively,
evaluated at the saddle point (sp). In this result, an integration over the saddle-point manifold
remains. The integration over the remaining space of sequences of transversal wave vectors was
evaluated in Gaussian approximation. Instead of the usual determinant of the Hessian matrix, we here
obtain its pseudo-determinant $\text{pdet}(\mathsf{H})$ which discards the zero eigenvalues associated with
the existence of a saddle-point manifold.

Before we continue with the evaluation of the leading saddle-point approximation, we
want to give an interpretation of the saddle point in terms of geometrical optics. 
According to \eqref{eq:saddle_point}, the transverse wave vector is conserved at the saddle
point, which means that the main contribution to the Casimir free energy comes from the reflection
at a plane perpendicular to the axis connecting the two sphere centers. Such planes are tangential
to the spheres at the points of closest distance between the spherical surfaces. Moreover, with the
transverse wave vector conserved, the scattering plane and the Fresnel plane coincide. As a consequence,
the coefficients in \eqref{eq:reflection_matrix_elements} reflecting the change of the
polarization basis simplify to $A=1$, $B=C=D=0$ at the saddle-point manifold. 

From \eqref{eq:def_f}, \eqref{eq:def_eta}, and the conservation of the transverse wave vector
\eqref{eq:saddle_point}, one finds $f_\SP=0$. Making use of the result
for the pseudo-determinant derived in Ref.~\onlinecite{Spreng2018}, one finds for the
leading-order saddle-point approximation \eqref{eq:trMr_LO-SPA}
\begin{equation}
\label{eq:trMr_SPA}
\left[\tr\mathcal{M}^r\right]_\text{LO-SPA} = \frac{R_\mathrm{eff}}{4\pi r}
\int \frac{\mathrm{d}\mathbf{k}_\SP}{\kappa_\SP} \kappa_\SP^{2r}
g_\SP
\end{equation}
with the effective radius 
\begin{equation}
R_\mathrm{eff} = \frac{R_1R_2}{R_1 + R_2}
\end{equation}
and
\begin{equation}
\label{eq:g_sp}
g_\SP = 
\frac{e^{-2r\kappa_\SP L}}{\kappa_\SP^{2r}}  \sum_{p_1,\ldots, p_{2r}} 
\prod_{j=1}^r 
\tilde{r}^{(1)}_{p_{2j+1} p_{2j}} 
\tilde{r}^{(2)}_{p_{2j} p_{2j-1}}\,.
\end{equation}
The function $g$ at the saddle point can easily be calculated for dielectric spheres
\cite{Spreng2021} because the polarization is conserved. For bi-isotropic spheres,
it is more convenient to first sum over all round-trips $r$ and then to evaluate the
sum over the polarizations as we will demonstrate now. This approach has already been
used successfully to determine the exact high-temperature limit of the Casimir free
energy for two Drude spheres of arbitrary size.\cite{Schoger2021}

To obtain the well-known PFA expression for the Casimir force, we need to take 
the negative derivative of the Casimir free energy with respect to the surface-to-surface
distance $L$. With \eqref{eq:roundtrip_decomposition} and \eqref{eq:trMr_SPA} the
contribution to the Casimir force from an imaginary frequency $\xi$ reads
\begin{equation}
\label{eq:F_xi}
 \mathbb{F}_\text{LO-SPA}(\xi) = -\frac{\partial\mathfrak{F}_\text{LO-SPA}(\xi)}{\partial L}
	              = -2\pi R_\mathrm{eff} \int \frac{\mathrm{d}\mathbf{k}_\SP}{(2\pi)^2} 
 \mathcal{P}(\mathbf{k}_\SP)\,,
\end{equation}
where we introduced 
\begin{equation}
\label{eq:P_of_kappa}
\mathcal{P}(\mathbf{k}_\SP) = \sum_{r=1}^\infty 
\frac{e^{-2r\kappa_\SP L}}{r}\sum_{p_1,\ldots, p_{2r}} \prod_{j=1}^r 
\tilde{r}^{(1)}_{p_{2j+1} p_{2j}} 
\tilde{r}^{(2)}_{p_{2j} p_{2j-1}}\,.
\end{equation}
While the polarizations at the beginning and the end of $r$ round-trips coincide because of
the trace, $\mathcal{P}$ accounts for all possible sequences of polarizations in-between.
In order to account for all polarization sequences, it is convenient to introduce a function
$h_r^{p'p}$ describing the contribution of $r$ round-trips starting with a mode of
polarization $p$ incident on sphere~2 and ending with a  mode of polarization $p'$ reflected
from sphere~1. The sequence of round-trips is closed by a translation from sphere~1 to sphere~2
where in the plane-wave basis no polarization change can occur.

We then write \eqref{eq:P_of_kappa} as
\begin{equation}
\label{eq:polarizations}
\mathcal{P} = \sum_{r=1}^\infty \frac{1}{r} (h_r^{\TM,\TM} + h_r^{\TE,\TE})\,.
\end{equation}
The function $h_r^{p'p}$ can be expressed recursively by decomposing $r$ round-trips into
a single round-trip where the polarization can either change or not, followed by the
remaining $r-1$ round-trips
\begin{align}
\label{eq:def_hrPP'}
h_r^{p'p} &= 
a^{p', \TM} h_{r-1}^{\TM, p} +  a^{p', \TE} h_{r-1}^{\TE, p}
\end{align}
with
\begin{equation}
\label{eq:def_APP'}
a^{p' p} = h_1^{p'p} = \left(\tilde{r}_{p', \TM}^{(1)} \tilde{r}_{\TM, p}^{(2)} +
\tilde{r}_{p', \TE}^{(1)} \tilde{r}_{\TE, p}^{(2)} \right)
e^{-2\kappa_\SP L}\,.
\end{equation}

The sum over round-trips can be conveniently dealt with by introducing a generating
function for $h_r^{p'p}$ as
\begin{equation}
\label{eq:defHPP'}
H^{p'p}(t)  = \sum_{r=1}^\infty t^{r}h_r^{p'p}\,,
\end{equation}
where the parameter $t$ keeps track of the number of round-trips. 
From the recursion relation \eqref{eq:def_hrPP'}, one can immediately derive a 
corresponding recursion relation for the generating function
\begin{align}
\label{eq:H_recursion}
H^{p'p}(t) &= ta^{p'p} + t\left(a^{p', \TM}H^{\TM, p} + a^{p', \TE}H^{\TE, p} \right)\,.
\end{align}
Making use of the generating function, the factor $1/r$ in \eqref{eq:polarizations}
can be written as a definite integral and we obtain
\begin{equation}
\label{eq:P_integral}
\mathcal{P} = \int_{0}^{1}  \frac{\mathrm{d} t}{t}
\left[H^{\TM,\TM}(t) + H^{\TE,\TE}(t)\right]\,.
\end{equation}
The set of recursion relations \eqref{eq:H_recursion} can be used to express the main part
of the integrand in terms of single round-trip contributions \eqref{eq:def_hrPP'} as
\begin{widetext}
\begin{equation}
H^{\TM,\TM}(t) + H^{\TE,\TE}(t)
= t\frac{a^{\TM,\TM} + a^{\TE,\TE} - 2t a^{\TM,\TM}a^{\TE,\TE} + 2t a^{\TM,\TE}a^{\TE,\TM}}
{(1- ta^{\TM,\TM})(1- ta^{\TE,\TE}) - t^2a^{\TM,\TE} a^{\TE,\TM}}\,.
\end{equation}
\end{widetext}
Defining a matrix $\mathsf{A}$ through the matrix elements \eqref{eq:def_APP'}, the denominator 
is given by the determinant of $\mathbb{I}-t\mathsf{A}$ while the numerator is given by its negative derivative with
respect to $t$. Therefore, it is straightforward to carry out the integral in \eqref{eq:P_integral}
and we arrive at
\begin{equation}
\label{eq:result_P}
\mathcal{P} = 
-\log\det\left(\mathbb{I} - \mathsf{A}\right)\,.
\end{equation}

As in this section we are interested only in the leading term for large sphere
radii, it is sufficient to keep the leading-order terms of the reflection
coefficients \eqref{eq:ref_coef_corr}. The matrix $\mathsf{A}_0$ with the index
indicating the leading order then corresponds to the round-trip matrix between
two planes at a distance $L$
\begin{equation}
\label{eq:def_A0}
\mathsf{A}_0 =  \mathsf{R}_1 \mathsf{R}_2 e^{-2\kappa_\SP L}
\end{equation}
with the reflection matrix at a planar surface 
\begin{equation}
\label{eq:reflection_matrix}
\mathsf{R}_i  = \left(\begin{array}{cc}
r^{(i)}_{\TM,\TM} & r^{(i)}_{\TM, \TE}  \\
r^{(i)}_{\TE, \TM} & r^{(i)}_{\TE ,\TE}
\end{array}
\right)\,.
\end{equation}
Inserting \eqref{eq:result_P} together with \eqref{eq:def_A0} into \eqref{eq:F_xi}, 
we obtain the Lifshitz formula 
\begin{equation}
\mathbb{F}_0(\xi)  = 2\pi \Reff\int \frac{\mathrm{d}\mathbf{k}_\SP}{(2\pi)^2}  
\log\det\left(\mathbb{I} - \mathsf{R}_1\mathsf{R}_{2} e^{-2\kappa_\SP L}\right)\,,
\end{equation}
where the index 0 still indicates that only the leading terms of the scattering
amplitudes were taken into account.

By means of an integration over all frequencies or the evaluation of a sum over 
Matsubara frequencies in analogy to the expressions \eqref{eq:casimir_energy_imag}
and \eqref{eq:casimir_free_energy} for the Casimir free energy, we obtain the
Casimir force at temperature $T$ between two spheres at closest surface-to-surface
distance $L$ within the proximity-force approximation
\begin{equation}
 F_\text{PFA}(L,T) = 2\pi\Reff\mathcal{F}_\text{PP}(L,T)\,,
\end{equation}
where $\mathcal{F}_\text{PP}(L, T)$ is the Casimir free energy per unit area between two planes
at distance $L$. As the reflection matrix \eqref{eq:reflection_matrix} needs not
be diagonal, this result is valid also for bi-isotropic media.

The Casimir free energy can be obtained by integrating the Casimir force with respect 
to the distance from $L$ to infinity. This integral can be carried out if the round-trip
matrix \eqref{eq:def_A0} is represented in its eigenbasis. We thus find 
\begin{equation}
\mathfrak{F}_\text{PFA}(\xi) = - \frac{R_\mathrm{eff}}{4\pi}
\int \frac{\mathrm{d}\mathbf{k}_\SP}{\kappa_\SP} \left[
\mathrm{Li}_2(\lambda_1) + \mathrm{Li}_2(\lambda_2) 
\right]\,,
\label{eq:F_PFA}
\end{equation}
where $\lambda_{1,2}$ are the eigenvalues of \eqref{eq:def_A0} and $\mathrm{Li}_n(\lambda)$
is the polylogarithm of order $n$.\cite{DLMF}

Here, we consider the special case of two PEMC spheres. Making use of the
leading terms in the reflections coefficients
\eqref{eq:rTETE_pemc}-\eqref{eq:rTETM_pemc}, the eigenvalues of $\mathsf{A}_0$ are given by
\begin{align}
\lambda_{1/2} = \exp(\pm 2i\delta)\exp(-2\kappa_\SP L)\,,
\end{align}
where we adopted the notation used in Ref.~\onlinecite{Rode2018} with
\begin{equation}
\label{eq:def_delta}
\delta = \theta^{(2)} - \theta^{(1)}
\end{equation}
taking values between $0$ and $\pi/2$. The lower bound corresponds to a setup
consisting of two identical spheres while for the upper bound one sphere is perfectly
conducting whereas the other sphere has infinite permeability. More explicit results
for PEMC spheres will be given in Sec.~\ref{sec:PEMC_T0} where the case of zero temperature
is addressed.

\section{Beyond PFA: Leading-order corrections}
\label{sec:corrections_to_pfa}

In the previous section, we have made use of two approximations to
derive PFA. We restricted ourselves to the leading term in the reflection
coefficients and in addition evaluated only the leading term of the saddle-point
approximation as is usually done. As a consequence, corrections arise from two
sources and can be attributed different physical meanings.

Taking the first subleading term of the reflection coefficients into account
amounts to going beyond geometrical optics and to allow for diffraction. This
contribution will be discussed in Sec.~\ref{sec:diffractive_corrections}.

For the second contribution, we remain within geometrical optics, \textit{i.e.}
we only keep the leading-order terms of the reflection coefficients, but go one
order further in the saddle-point approximation as explained in
Sec.~\ref{sec:geometric_corrections}. Then, the tangential plane at which
reflection takes place need no longer be perpendicular to the axis connecting
the two spheres.

As discussed at the end of Sec.~\ref{eq:refl_coeff_zero}, the reflection matrix
elements at zero frequency can be obtained from the finite-frequency
expressions in the limit of very large spheres. The PFA result given in the
previous section thus holds for arbitrary temperatures. However, this is no
longer the case for the leading corrections. Here, the contribution of the
zero-frequency term has to be calculated separately as was shown in detail in
Ref.~\onlinecite{Henning2021} for a perfectly reflecting sphere and plate. In the following, 
we will focus on the correction to the PFA result in the
zero-temperature limit based on the expression \eqref{eq:WKB_approximation}
for the reflection matrix elements at non-zero imaginary frequencies. 

While the results for the two correction terms derived in
Secs.~\ref{sec:diffractive_corrections} and \ref{sec:geometric_corrections} are
rather general, restricting ourselves to PEMC materials and zero temperature in
Sec.~\ref{sec:PEMC_T0} will allow us to give explicit results for the leading
correction to the Casimir energy and to discuss its dependence on a material
parameter.

\subsection{Diffractive corrections to PFA} 
\label{sec:diffractive_corrections}

The corrections from the leading saddle-point approximation can be obtained by replacing the 
matrix $\mathsf{A}$ in \eqref{eq:result_P} with 
\begin{equation}
\mathsf{A} = \mathsf{A}_0 + \frac{c}{\mathcal{\xi}} \mathsf{A}_1\,,
\end{equation} 
where $\mathsf{A_0}$ is given by \eqref{eq:def_A0} and the matrix $\mathsf{A}_1$
takes the leading corrections of the scattering amplitudes due to diffraction into 
account. According to \eqref{eq:ref_coef_corr}, the matrix elements of $\mathsf{A}_1$
are given by
\begin{equation}
\label{eq:def_a1}
a^{p'p}_1 = \sum_{q= \TM, \TE} r_{p' q}^{(1)} r_{q p}^{(2)} 
\left(\frac{s_{p' q}^{(1)} }{R_1}+\frac{s_{q p}^{(2)} }{R_2}\right)
e^{-2\kappa_\SP L}\,.
\end{equation}
Expanding the logarithm in \eqref{eq:result_P} one finds up to the leading correction
\begin{equation}
\label{eq:P_with_corr}
\mathcal{P} = -
	\log\det\left(\mathbb{I} - \mathsf{A}_0\right)
	+ \frac{c}{\xi} \tr\left[(\mathbb{I}-\mathsf{A}_0)^{-1} \mathsf{A}_1\right]\,.
\end{equation}
With the eigenvalues $\lambda_{1,2}$ of the round trip matrix $\mathsf{A}_0$, the
trace can be expressed as 
\begin{equation}
\tr\left[(\mathbb{I}-\mathsf{A}_0)^{-1} \mathsf{A}_1\right]
= \frac{\alpha_0 + \alpha_1}{(1-\lambda_1)(1-\lambda_2)}\,,
\end{equation}
where the expansion coefficients $\alpha_{0,1}$ are given by
\begin{equation}
\label{eq:expansion_coef}
\begin{aligned}
\alpha_0 &= \tr\mathsf{A}_1 = a^{\TM, \TM}_1 + a^{\TE, \TE}_1\,, \\
\alpha_1 &= - a^{\TM, \TM}_0 a^{\TE, \TE}_1 -  a^{\TE, \TE}_0 a^{\TM, \TM}_1 \\
	&\quad + a^{\TM, \TE}_0 a^{\TE, \TM}_1 +  a^{\TE, \TM}_0 a^{\TM, \TE}_1\,. 
\end{aligned}
\end{equation}

Now, we can obtain the Casimir force including the leading-order correction by inserting
\eqref{eq:P_with_corr} into the expression on the right-hand side of \eqref{eq:F_xi}.
Here, we are interested in the Casimir free energy and integrate over the surface-to-surface
distance from $L$ to infinity. The contribution arising from the leading-order saddle-point
approximation including the leading diffractive correction
\begin{equation}
\mathfrak{F}_\text{LO-SPA}(\xi) = \mathfrak{F}_\text{PFA}(\xi) + \mathfrak{F}_\text{diff}(\xi)
\end{equation}
then contains the PFA expression \eqref{eq:F_PFA} as well as the leading diffractive
correction
\begin{widetext}
\begin{equation}
\mathfrak{F}_\text{diff}(\xi)
= - \frac{cR_\mathrm{eff}}{2\pi\xi} \int \mathrm{d}\mathbf{k}_\SP
\int_{L}^\infty \mathrm{d}l 
\frac{\left(\alpha_0 + \alpha_1 e^{-2\kappa_\SP(l-L)}\right) e^{-2\kappa_\SP(l-L)}}
{\left(1- \lambda_1e^{-2\kappa_\SP(l-L)}\right)\left(1- \lambda_2 e^{-2\kappa_\SP(l-L)}\right)}\,,
\end{equation}
where the exponential factors arise from the length dependence in \eqref{eq:def_A0} and
\eqref{eq:def_a1}. Evaluating the integral over $l$, one finally arrives at
\begin{equation}
\mathfrak{F}_\text{diff}(\xi) = \frac{cR_\mathrm{eff}}{4\pi\xi}
\int \frac{\mathrm{d}\mathbf{k}_\SP}{\kappa_\SP} \Bigg\{
 \frac{1}{\lambda_1 - \lambda_2}
 \Big[(\alpha_0 + \frac{\alpha_1}{\lambda_1})\log(1-\lambda_1)
	- (\alpha_0 + \frac{\alpha_1}{\lambda_2})\log(1-\lambda_2)
\Big]
\Bigg\}\,.
\label{eq:F_diff}
\end{equation}
\end{widetext}

This expression for the Casimir free energy simplifies considerably for PEMC spheres. 
Making use of the expressions \eqref{eq:rTETE_pemc}--\eqref{eq:rTETM_pemc} for the reflection
coefficients, the expansion coefficients \eqref{eq:expansion_coef} yield 
$\alpha_0 = -\cos(2\delta)\alpha_1$ with $\alpha_1 = \xi/c\Reff \kappa_\SP$ and 
the diffractive correction \eqref{eq:F_diff} is thus given by
\begin{equation}
\label{eq:diff_pemc}
\mathfrak{F}_\text{diff}(\xi) = -\frac{1}{4} \int\frac{\mathrm{d}\mathbf{k}_\mathrm{sp}}{2\pi\kappa^2_\mathrm{sp}}
[\log(1-\lambda_1) + \log(1-\lambda_2)]\,.
\end{equation}

\subsection{Geometrical corrections to PFA}
\label{sec:geometric_corrections}

We now turn to the second source of leading corrections where the reflection matrix elements
are still given by the leading term implying the use of geometrical optics but where we 
consider the next-to-leading term in the saddle-point approximation (NTLO-SPA). For the trace
over the $r$-th power of the round-trip matrix in the notation employed in \eqref{eq:trMr_WKB}
one finds
\begin{multline}
\label{eq:trMr_NTLO-SPA}
\left[\tr\mathcal{M}^r\right]_\text{NTLO-SPA}
= \frac{R_\mathrm{eff}}{4r}\int \frac{\mathrm{d}\mathbf{k}_\SP}{2\pi\kappa_\SP} \kappa_\SP^{2r}
	\Bigg[g_{ij} \mathsf{H}^{ij} \\
	- f_{ijk}g_l \mathsf{H}^{ij} \mathsf{H}^{kl}
	-\frac{1}{4}g_\mathrm{sp} f_{ijkl} \mathsf{H}^{ij}\mathsf{H}^{kl} \\
	+\frac{1}{12}g_\mathrm{sp} f_{ijk}f_{lmn} 
	\left(3 \mathsf{H}^{ij}\mathsf{H}^{kl}\mathsf{H}^{mn} 
	      + 2\mathsf{H}^{il} \mathsf{H}^{jm}\mathsf{H}^{kn}\right)	
\Bigg]\,.
\end{multline}
For more details on the derivation, we refer the reader to appendix~B of
Ref.~\onlinecite{Henning2021}.  A lower index $i$ represents a derivative with
respect to $k_{i,\alpha}$ with $\alpha=x,y$ evaluated at the saddle point.
Upper indices specify matrix elements of the inverse Hessian matrix after the
direction of the saddle-point manifold associated with a zero eigenvalue has
been removed (cf.\ discussion related to \eqref{eq:trMr_LO-SPA}). It is
implicitly understood that a sum is taken over equal lower and upper indices.

As diffractive corrections are irrelevant here, only the leading-order term of
$g$ has to be taken into account. It was shown in Ref.~\onlinecite{Spreng2021} that
the fourth term in \eqref{eq:trMr_NTLO-SPA} vanishes due to the symmetry of the
function $f$ in \eqref{eq:def_f} with respect to its arguments. Furthermore,
the second term is found to vanish for dielectric spheres \cite{Spreng2021} and
for PEMC spheres.

Restricting ourselves to PEMC materials, the remaining terms yield \cite{Spreng2021}
\begin{multline}
\label{eq:NTLO_integrand}
g_{ij} \mathsf{H}^{ij}  - 
\frac{1}{4}g_\mathrm{sp} f_{ijkl} \mathsf{H}^{ij}\mathsf{H}^{kl} 
	+\frac{1}{6}g_\mathrm{sp} f_{ijk}f_{lmn} \mathsf{H}^{il} \mathsf{H}^{jm}\mathsf{H}^{kn} \\
 = - g_\SP \frac{rL\kappa_\SP (c^2\kappa_\SP^2 + \xi^2) + \xi^2}{3 r c^2 \kappa_\SP^3 \Reff}
	 (r^2 + 3u -1)\,,
\end{multline}
where $g_\SP$ depends on the material parameter $\delta$ introduced in \eqref{eq:def_delta}.
The parameter
\begin{equation}
\label{eq:def_u}
u = \frac{R_1R_2}{(R_1+R_2)^2}
\end{equation}
characterizes the relative sphere radii with values between 0 corresponding to the plane-sphere geometry
and 1/4 corresponding to equal radii.

Inserting \eqref{eq:NTLO_integrand} into \eqref{eq:trMr_NTLO-SPA} and
evaluating the sum of round-trips in \eqref{eq:roundtrip_decomposition} as well
as over the polarizations in $g_\SP$ as explained in
Sec.~\ref{sec:asymptotic_expansion} one finds for the geometrical correction to
the contribution of the Casimir free energy at a given imaginary frequency
\begin{widetext}
\begin{equation}
\label{eq:geo_pemc}
\mathfrak{F}_\mathrm{geo}(\xi) = \frac{1}{12}\sum_{i = 1, 2}
\int \frac{\mathrm{d}\mathbf{k}_\SP}{2\pi\kappa_\SP^2}
  \Bigg\{ L \kappa_\SP \left(1 + \frac{\xi^2}{c^2\kappa_\SP^2}\right)
  \Big[\frac{\lambda_i}{1-\lambda_i} + (3u-1)\mathrm{Li}_{2}(\lambda_i) \Big] 
+ \frac{\xi^2}{c^2\kappa_\SP^2}\Big[-\log(1-\lambda_i) + (3u-1)\mathrm{Li}_{3}(\lambda_i) \Big] 
\Bigg\}\,.
\end{equation}
\end{widetext}

For PEMC spheres, we can now add \eqref{eq:F_PFA}, \eqref{eq:diff_pemc}, and \eqref{eq:geo_pemc}
to obtain
\begin{equation}
 \mathfrak{F}(\xi) = \mathfrak{F}_\text{PFA}(\xi) + \mathfrak{F}_\text{diff}(\xi) +
	            \mathfrak{F}_\text{geo}(\xi)\,.
\end{equation}
This result will be the basis for an explicit evaluation of the Casimir energy at zero temperature
for PEMC spheres in the next section.

\subsection{PEMC spheres at zero temperature}
\label{sec:PEMC_T0}

The Casimir energy is obtained by integrating $\mathfrak{F}(\xi)$ according to 
\eqref{eq:casimir_energy_imag}. Taking leading corrections into account, the
Casimir energy is usually expressed as
\begin{equation}
\label{eq:E_expansion}
E = E_\mathrm{PFA}\left[1 +\beta_1 x + o\left(x\right)\right]\,,
\end{equation}
where 
\begin{equation}
\label{eq:def_x}
x = \frac{L}{\Reff}
\end{equation}
characterizes the dimensionless distance between the spheres and $o(x)$ denotes a contribution going
faster to zero than $x$. The coefficient
\begin{equation}
\label{eq:beta1_decomposition}
\beta_1 = \beta_\text{diff} + \beta_\text{geo}
\end{equation}
accounts for the two contributions discussed in Sec.~\ref{sec:diffractive_corrections} and
\ref{sec:geometric_corrections}, respectively.

The PFA result for two PEMC spheres is obtained by carrying out the integral in \eqref{eq:F_PFA}
and over the imaginary frequency. The result can be simplified by means of the Jonquière inversion
formula for the sum of the polylogarithms \cite{Jonq1889}
\begin{align}
\label{eq:jonquiere}
\mathrm{Li}_n[e^{2\pi iz}]  + (-1)^n \mathrm{Li}_n[e^{-2\pi iz}] = - \frac{(2\pi i)^n}{n!}B_n(z)\,,
\end{align}
where $B_n(z)$ are Bernoulli polynomials. One finally obtains
\begin{equation}
\label{eq:EPFA_PEMC}
E_\mathrm{PFA} = - \frac{\hbar c R_\mathrm{eff}}{720\pi L^2}
	\left[\pi^4-30\delta^2(\pi-\delta)^2\right]\,,
\end{equation}
which is consistent with the Casimir energy for parallel PEMC plates obtained in
Ref.~\onlinecite{Rode2018}. 

After evaluation of the integrals in \eqref{eq:casimir_energy_imag}, \eqref{eq:diff_pemc},
and \eqref{eq:geo_pemc} one obtains with the help of \eqref{eq:jonquiere} for the two
contributions to the coefficient \eqref{eq:beta1_decomposition}
\begin{equation}
\label{eq:beta_diff_geo}
\beta_\mathrm{diff} = -15\frac{\pi^2 - 6\delta(\pi-\delta)}{\pi^4 - 30\delta^2(\pi-\delta)^2}\,,\qquad
\beta_\text{geo} = \frac{1}{3} - u + \frac{\beta_\text{diff}}{3}\,.
\end{equation}
The leading correction for the Casimir energy, \text{i.e.} the term in \eqref{eq:E_expansion}
depending on $\beta_1$, thus becomes
\begin{equation}
 \begin{aligned}
 \label{eq:def_e1}
  E_1 &= \frac{\hbar c}{720\pi L}\bigg[20\big(\pi^2-6\delta(\pi-\delta)\big)\\
      &\qquad\qquad\quad  -\bigg(\frac{1}{3}-u\bigg)\big(\pi^4-30\delta^2(\pi-\delta)^2\big)\bigg]\,.
 \end{aligned}
\end{equation}

For $\delta = 0$, we reproduce the known results for two perfect electric conductors
\begin{equation}
\label{eq:beta_perfect_reflector}
E_\text{PFA}^\text{PEC} = -\frac{\hbar c \pi^3 \Reff}{720 L^2}\,,\qquad
\beta_1^\text{PEC} = \frac{1}{3} - \frac{20}{\pi^2}-u\,.
\end{equation}
On the other hand, $\delta = \pi/2$ corresponds to the Boyer setup \cite{Boyer1974} with a
perfectly conducting sphere and one with infinite permeability, for which we obtain
\begin{equation}
\label{eq:beta_Boyer}
E_\text{PFA}^\text{Boyer} = \frac{7\hbar c \pi^3 \Reff}{5760 L^2}\,, \qquad
\beta_1^\text{Boyer} = \frac{1}{3} - \frac{80}{7\pi^2}-u\,.
\end{equation}
These limiting cases for two PEMC spheres are known in the literature \cite{Teo2012, Bimonte2012}
where it has been noticed that the first two leading terms can be obtained as sum of the corresponding
terms for two scalar fields. The Casimir energy for two perfect mirrors equals the sum of the energies
for two Dirichlet spheres and for two Neumann spheres. For the Boyer setup, on the other hand, the Casimir
energy is equivalent to the case where one sphere obeys Dirichlet boundary conditions while the other
one obeys Neumann boundary conditions. In fact, it has been stated in Ref.~\onlinecite{Bimonte2012}
that the former case holds for any perfect conductors as long as the interacting objects are smooth.

Comparing \eqref{eq:beta_perfect_reflector} and \eqref{eq:beta_Boyer}, one
notices that the Casimir energy within PFA changes sign and thus becomes zero
at a certain value of $\delta_\text{crit}$. This behavior has been discussed in
Ref.~\onlinecite{Rode2018} for parallel PEMC plates and is already known for plates
with pseudo-periodic boundary conditions \cite{Asorey2013}. For a sphere-sphere
setup, the Casimir energy \eqref{eq:EPFA_PEMC} within PFA then will also vanish
for $\delta_\text{crit}$. In contrast, in the first correction \eqref{eq:def_e1}
only the second term in the square brackets vanishes at $\delta_\text{crit}$ while
the first term yields a non-vanishing result.

\begin{figure}
\centering
\includegraphics[width=0.9\columnwidth]{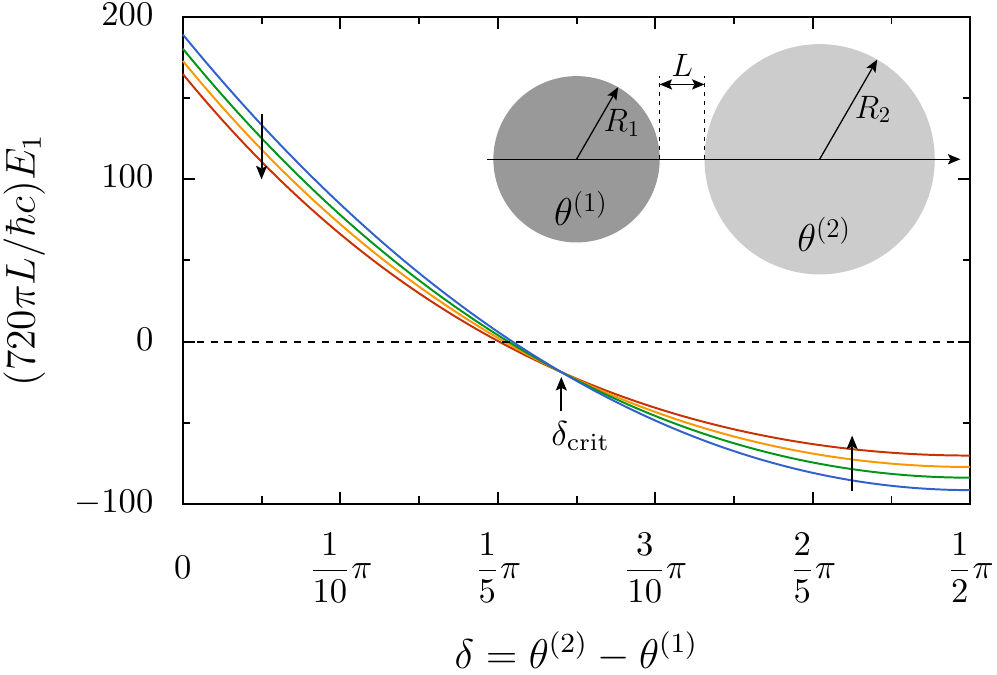}
\caption{Lowest-order correction $E_1$ of the Casimir energy for PEMC spheres
         characterized by the material parameters $\theta^{(1)}$ and $\theta^{(2)}$
	 as function of $\delta=\theta^{(2)}-\theta^{(1)}$. The corrections are
	 plotted for the sphere-sphere geometry with ratios $R_1/R_2 = 1$ (blue),
	 4 (green) and 10 (yellow) as well as for the limiting case of the plane-sphere geometry
	 where $R_1/R_2 = \infty$ (red). The direction of increasing ratios is marked
	 by two arrows. The PFA result vanishes for $\delta_\text{crit}$, making the correction
	 $E_1$ the leading-order term of the Casimir energy.}
\label{fig:pemc}
\end{figure}

The dependence of the leading correction $E_1$ to the Casimir energy is
displayed in Fig.~\ref{fig:pemc} as a function of $\delta$ and for various
values of $R_1/R_2$ including the case of the plane-sphere geometry. At
$\delta_\text{crit}$, the leading-order correction becomes actually the leading
contribution and around this critical value, it can dominate the Casimir energy
even for rather small distances between the spheres.  Fig.~\ref{fig:pemc} also
shows a relatively weak dependence on the ratio of sphere radii. This
dependence vanishes at $\delta_\text{crit}$ because the first term in
\eqref{eq:def_e1} does not depend on $u$. 

\section{Next-to-leading order corrections for perfectly reflecting spheres at zero temperature}
\label{sec:NTLO_correction}

\begin{figure}
 \begin{center}
  \includegraphics[width=0.9\columnwidth]{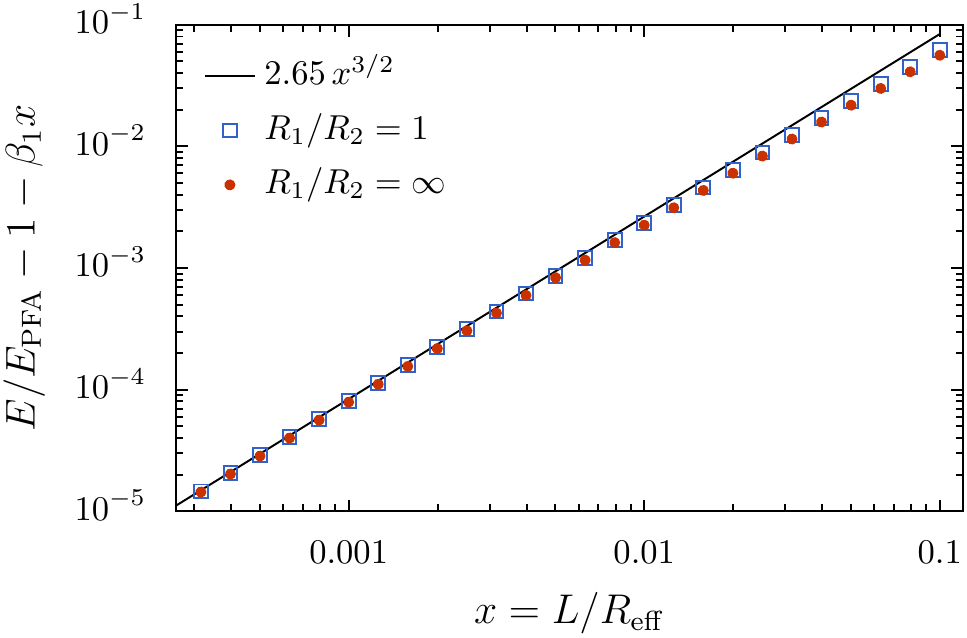}
 \end{center}
 \caption{Numerical results for the correction to the Casimir energy beyond the linear term
	  are shown for perfectly reflecting spheres of equal radii (open squares) and
	  a plane-sphere setup (filled circles) at separation $L$ in vacuum at zero
	  temperature. The solid line, representing $2.65\,x^{3/2}$, corresponds to
	  a numerical fit on the plane-sphere results obtained in Ref.~\onlinecite{Hartmann2018b}.}
 \label{fig:ntlo_correction}
\end{figure}

Finally, we will explore corrections beyond the leading ones discussed in the two previous
sections. For simplicity, we will restrict ourselves to two perfectly reflecting spheres,
\textit{i.e.} spheres consisting of perfect electric conductors, at zero temperature. Anticipating
the numerical results shown in Fig.~\ref{fig:ntlo_correction}, we write the asymptotic expansion
of the Casimir energy as
\begin{equation}\label{eq:energy_asymptotics}
	E = E_\text{PFA}\left[ 1 + \beta_1 x + \beta_{3/2} x^{3/2} + \ldots \right]
\end{equation}
with the aspect ratio defined in \eqref{eq:def_x} and the PFA result for
perfect electric conductors given in \eqref{eq:beta_perfect_reflector}. As was
found for $\beta_\text{geo}$ in \eqref{eq:beta_diff_geo}, the coefficient
$\beta_{3/2}$ may in general depend on the ratio of the sphere radii through
the dimensionless quantity $u$ defined in \eqref{eq:def_u}.

In Fig.~\ref{fig:ntlo_correction}, numerical results for the correction beyond
the linear term are shown for two spheres of equal radii (open squares) and
a plane-sphere setup (filled circles). Results for spheres with different radii
have been found to lie between these two data sets. The solid line represents the
function $2.65\,x^{3/2}$ obtained by a numerical fit to the plane-sphere
results at $x\leq10^{-3}$ \cite{Hartmann2018b}. The data for the sphere-sphere geometry appear to
approach the same asymptotic behavior for the next-to-leading-order (NTLO)
correction. With increasing aspect ratio $x$, the data tend to deviate from the
fit because of higher order corrections.

On the basis of our considerations in previous sections, one might have
expected that already the NTLO correction should be proportional to $x^2$. The
NTLO correction to the saddle-point approximation would give rise to such a
term as would the NTLO correction of the Mie scattering amplitudes $S_{\TE,\TE}$ and
$S_{\TM, \TM}$ for perfect reflectors (cf.\ Fig. 7.5 on page 99 in Ref.~\onlinecite{Spreng2021}).  However, this
expectation is clearly refuted by the numerical results. Our aim in the
remainder of this section is thus to understand the origin of a correction
proportional to $x^{3/2}$. We will refrain from trying to obtain the prefactor
analytically which would require the push the evaluation of the saddle-point
approximation even one order further than we did in
Sec.~\ref{sec:geometric_corrections}. Instead, we will show that already an
appropriate evaluation of diffractive corrections will yield the observed power
law.

Proceeding as in Sec.~\ref{sec:asymptotic_expansion}, one would encounter
a logarithmic divergence in the round-trip sum. Instead we follow a strategy
similar to the one employed in Ref.~\onlinecite{Henning2021} where the corrections to
PFA were calculated in an intermediate temperature regime. For perfectly reflecting
spheres, only the polarization preserving scattering amplitudes
\eqref{eq:WKB_scattering_amplitude} are nonvanishing and the diffractive correction
$s_{pp}$ is given by the first term in \eqref{eq:sTETE} and \eqref{eq:sTMTM} for
$p=\TE,\TM$, respectively. The key ingredient for the calculation of the NTLO
correction consists now in replacing $1 + s_{pp}/\tilde{\xi}$ by $\exp(s_{pp}/\tilde\xi)$
in \eqref{eq:ref_coef_corr} which is correct up to order $\tilde\xi^{-1}$. Note that
the leading corrections $s_{pp}$ of the scattering amplitudes take on negative values, 
thus ensuring later the convergence of the round-trip sum.

This exponential replacement implies that diffractive corrections are taken into account for 
an arbitrary number of reflections during the round-trips. In contrast, in
Sec.~\ref{sec:diffractive_corrections} a diffractive correction was taken
into account only at one of the reflections during the series of round-trips. 
Mathematically, the exponential replacement amounts to a resummation of higher-order
diffractive corrections. 

Applying the exponential replacement in \eqref{eq:trMr_SPA}, we can express the
leading-order of the saddle-point approximation as
\begin{widetext}
\begin{equation}\label{eq:NTLO-SPA_trace}
\left[\tr\mathcal{M}^r\right]_\text{LO-SPA} = \frac{\Reff}{2r} \sum_p \int_0^\infty dk_\SP
\frac{k_\SP}{\kappa_\SP}\exp\left[-r\left(2\kappa_\SP L - \frac{c}{\xi\Reff}s_{pp}\right)\right]\,,
\end{equation}
where it is implicitly understood that the functions $s_{pp}$ have been evaluated on the
saddle-point manifold \eqref{eq:saddle_point}.

An expression for the Casimir energy can now be obtained from \eqref{eq:NTLO-SPA_trace}
together with \eqref{eq:casimir_energy_imag} and \eqref{eq:roundtrip_decomposition}.
It is convenient to introduce a new integration variable $t=\xi/c\kappa_\SP$ to obtain
\begin{equation}\label{eq:E_TE}
E_\text{LO-SPA} = - \frac{\hbar c \Reff}{4\pi} \sum_p \int_0^1 dt \sum_{r=1}^\infty \frac{1}{r^2} 
\int_0^\infty d\kappa_\SP\,\kappa_\SP\exp\left[-2r\left(\kappa_\SP L 
	+ \frac{1}{\kappa_\SP \Reff}\sigma_p\right)\right]\,,
\end{equation}
\end{widetext}
where
\begin{equation}
\begin{aligned}
\sigma_\TE &\equiv -\left.\frac{c\kappa}{2\xi}s_{\TE,\TE}\right\vert_\SP = \frac{2-t^2}{4}\,,\\ 
\sigma_\TM &\equiv -\left.\frac{c\kappa}{2\xi} s_{\TM,\TM}\right\vert_\SP = \frac{t^2}{4}
\end{aligned}
\end{equation}
are positive in the range of integration. The integration over $\kappa_\SP$ can now
be carried out yielding essentially a modified Bessel function of the second kind
$K_2(4r\sqrt{\sigma_px})$.\cite{GradshteynRyzhik1980}

We now need to find an asymptotic expansion for $x\ll 1$. Indeed an asymptotic expansion
of the sum over round-trips containing the modified Bessel function can be worked out
using a method from Ref.~\onlinecite{Paris2018}. We find
\begin{equation}
\sum_{r=1}^\infty \frac{K_2(4 r \sqrt{z})}{r^2} \sim \frac{\pi^4}{720 z} - \frac{\pi^2}{12} + \frac{2\pi\sqrt{z}}{3} + O(z\log(z))
\end{equation}
for $z\ll 1$. Evaluating finally the integral over $t$, the Casimir energy in lowest
order of the saddle-point approximation, but including diffractive corrections yields
\begin{equation}
 \begin{aligned}
  E_\text{LO-SPA} &= -\frac{\hbar c \pi^3\Reff}{720 L^2}\bigg[1 - \frac{15}{\pi^2}x\\
	          &\qquad\qquad\qquad+ \frac{15(10+3\pi)}{4\pi^3}x^{3/2}+\dots\bigg]\,.
 \end{aligned}
\end{equation}
As expected, this result reproduces the PFA result \eqref{eq:beta_perfect_reflector}
and the leading-order correction due to diffraction as given by $\beta_\text{diff}$
in \eqref{eq:beta_diff_geo} for $\delta=0$. The NTLO correction is indeed found to
go with the power 3/2 of the aspect ratio $x$ but the prefactor accounts only for 
about 89\% of the numerical result. This discrepancy can be explained by the fact
that the NTLO-SPA and NNTLO-SPA contributions have been neglected in the calculation.
Since the diffractive correction is independent of $u$, only a small contribution from
other corrections can depend on $u$, thus explaining the weak dependence of
$\beta_{3/2}$ on $u$ found in the numerical results. Moreover, analyzing the contributions
of the individual polarizations, one finds that the main contribution comes from TE
polarization with about $90\%$ of the full correction.

Finally, we note that our calculation can be straightforwardly extended to real
dielectric materials, implying that the appearance of the $x^{3/2}$ term in the
asymptotic expansion should not be restricted to perfectly reflecting spheres.
Numerical results for the plane-sphere geometry indeed show that within the Drude
or the plasma model describing the metallic objects, a $x^{3/2}$-term does occur
\cite{Hartmann2018b}.

\section{Conclusions}\label{sec:conclusions}

Plane waves can constitute a basis well suited for the study of the Casimir
effect as we have shown here by reviewing recent work in this direction. They
turn out to be useful for numerical as well as analytical work, in particular
when the distance between the involved objects is small as is the case in most
experiments.  As we demonstrated, plane waves lend themselves particularly
well for an interpretation in terms of geometrical optics and diffractive
corrections.

For a setup consisting of two spheres with an arbitrary ratio of radii in
vacuum, we have demonstrated that the proximity-force approximation can be
obtained as the leading term in an asymptotic expansion for large radii. 
A previous calculation based on the saddle-point approximation of the trace
over a given number of round-trips of electromagnetic waves between the spheres
was extended to spheres made of bi-isotropic material where one needs to account
for polarization mixing during the reflection processes. The result was shown
to be naturally explained in terms of geometrical optics.

The framework provided by the saddle-point approximation allowed us to derive
leading-order corrections, both of geometrical and diffractive origin. Explicit
results were given for the first time for PEMC spheres at zero temperature.
It turned out that for a suitable choice of material parameters, the contribution
of the proximity-force approximation vanishes and the leading-order correction
actually becomes the dominant term in the Casimir energy.

Finally, numerical results for two perfectly reflecting spheres at small values of
$x=L/R_\mathrm{eff}$ motivated us to go one order further.
We found that the next-to-leading-order correction to PFA goes as  $x^{3/2}$ rather than $x^2.$
We discussed its origin as a resummation of higher-order diffractive corrections. 
The $x^{3/2}$ term implies that the estimation of the total correction to PFA based on the 
 the linear order alone is of limited interest for practical purposes, particularly in situations where beyond-PFA results are required. 

It can be expected that the usefulness of the plane-wave basis is not limited
to spherical objects but that this basis has potential for a study of the
Casimir interaction between a much wider class of systems.

\section*{Acknowledgments}

The authors would like to thank Michael Hartmann and Vinicius Henning for a
productive and enjoyable collaboration on various aspects of the material
reviewed here. Some of the results presented were obtained within a PROBRAL
collaboration supported by CAPES and DAAD. P.A.M.N. acknowledges financial
support by the Brazilian agencies National Council for Scientific and
Technological Development (CNPq), Coordination for the Improvement of Higher
Education Personnel (CAPES), the National Institute of Science and Technology
Complex Fluids (INCT-FCx), and the Research Foundations of the States of Rio de
Janeiro (FAPERJ) and São Paulo (FAPESP).

\bibliographystyle{apsrev}
\bibliography{casimir}

\end{document}